# From Morality Installation in LLMs to LLMs in Morality-as-a-System

Gunter Bombaerts
Eindhoven University of Technology, The Netherlands
g.bombaerts@tue.nl

## Abstract

Work on morality in large language models (LLMs) has progressed via constitutional AI, reinforcement learning from human feedback (RLHF) and systematic benchmarking, yet it still lacks tools to connect internal moral representations to regulatory obligations, to design cultural plurality across the full development stack, and to monitor how moral properties drift over the lifecycle of a deployed system. These difficulties reflect a shared root. Morality is installed in a model at training time. I propose instead a morality-as-a-system framework, grounded in Niklas Luhmann's social systems theory, that treats LLM morality as a dynamic, emergent property of a sociotechnical system. Moral behaviour in a deployed LLM is not fixed at training. It is continuously reproduced through interactions among seven structurally coupled components spanning the neural substrate, training data, alignment procedures, system prompts, moderation, runtime dynamics, and user interface.

This is a conceptual framework paper, not an empirical study. It philosophically reframes three known challenges, the interpretability-governance gap, the cross-component plurality problem, and the absence of lifecycle monitoring, as structural coupling failures that the installation paradigm cannot diagnose. For technical researchers, it explores three illustrative hypotheses about cross-component representational inconsistency, representation-level drift as an early safety signal, and the governance advantage of lifecycle monitoring. For philosophers and governance specialists, it offers a vocabulary for specifying substrate-level monitoring obligations within existing governance frameworks. The morality-as-a-system framework does not displace elements such as constitutional AI or RLHF it embeds them within a larger temporal and structural account and specifies the additional infrastructure those methods require.

**Keywords:** morality, Luhmann, large language models, sociotechnical systems, alignment, mechanistic interpretability, lifecycle governance, cultural plurality

***Declaration of AI Use.*** *The author used the free version of ChatGPT to assist in the preparation of this manuscript, including the generation and refinement of parts of the philosophical and technical content. All substantive arguments, interpretations, and conclusions remain the responsibility of the author. The author has reviewed the manuscript and assumes full responsibility for its content, accuracy, and integrity.*

# 1. Introduction

The dominant approach to morality in large language models treats moral behaviour as something to be installed: embedded via constitutional principles [1], fixed through RLHF [2], and verified through benchmarking [3]. This approach — which I call the *morality installation paradigm* — has produced genuine achievements. Constitutional AI produces models that reliably decline harmful requests. Benchmark suites such as ETHICS [3], MACHIAVELLI [4], and Delphi [5] enable systematic moral assessment. Mechanistic interpretability has begun identifying candidate evaluative structures within transformer representations [6], [7]. These are real contributions, and any alternative framework must build on them.

Yet structural challenges persist, of which Section 2 mentions three. Mechanistic interpretability and governance frameworks operate at different abstraction levels with no standardised protocol connecting them: safety engineers cannot currently verify whether internal representations satisfy regulatory requirements. Cultural plurality is managed at single training stages, leaving cross-stage compounding of cultural homogenisation unmodelled at the representation level. And while governance frameworks increasingly call for continuous monitoring, no established representation-level moral monitoring framework exists for tracking how moral properties change across fine-tuning and redeployment.

I argue these challenges share a common conceptual root: morality has been treated as a plug-in for LLMs rather than as a dynamic property of the broader sociotechnical system in which LLMs are embedded. The shift I propose is Copernican — shifting from the earth to the sun as center — in structure: rather than asking what moral values are installed in this LLM, I ask how morality operates as a system in which LLMs participate — moving the model from the centre to one component among several.

## 1.1. Audience and Contribution Type

**This is a conceptual framework paper with illustrative hypotheses, not an empirical study or formal-technical contribution.** It makes no new empirical claims and proposes no algorithms. What it offers is a conceptual vocabulary (Section 3) and an analytical architecture (Section 4).

**Technical researchers** (mechanistic interpretability, alignment, governance engineering) will find: a framing that recasts familiar technical challenges as structural coupling failures; three illustrative hypotheses specifying what the framework predicts, offered as philosophical scaffolding to guide experimental design rather than as technical specifications; and a seven-component decomposition of deployed LLM systems that maps onto actual team structures and supports audit design. The paper does not run experiments, does not provide formal guarantees, and does not specify measurement protocols at the level of production-ready tooling.

**Philosophers and AI ethicists** unfamiliar with Luhmann will find: a brief but accessible introduction to social systems theory applied to LLMs; an argument that systems thinking is relational rather than relativist — it does not dissolve moral standards but reconceives how they are produced and maintained; and a reframing of AI governance questions in terms that connect to existing sociotechnical and value-sensitive design literature. What the paper does not do: it does not resolve normative debates about which values should be encoded, does not replace ethical theory with systems description, and does not solve existing debates in system theory.

## 1.2. Contributions

**Contribution 1 (Conceptual architecture, Section 5):** A deployed LLM is better understood as a node in a larger moral sociotechnical system. I propose a seven-component provisional model of that system, with each component characterised by its own moral program and its structural coupling to others.

**Contribution 2 (Structural diagnosis, Section 6):** The morality-as-a-system reframing reveals why the three challenges are structural features of the installation paradigm rather than contingent research gaps. Each corresponds to a missing or underdeveloped structural coupling.

**Contribution 3 (Illustrative research agenda, Section 7):** The reframing generates three illustrative hypotheses and concrete research directions: linking interpretability to governance via representation-level lifecycle metrics, designing cross-component cultural plurality protocols, and extending auditing frameworks to include a mechanistic layer. These sketch, at the level of experimental design, not technical specifications, what technical researchers could do next.

# 2. The Morality Installation Paradigm and its Structural Challenges

The morality installation paradigm treats morality as a plug-in — a property to be installed, verified, and maintained at training time. Within this paradigm, central questions include: whose values should the model reflect? Can it reason morally rather than pattern-match on moral language? How should it handle cross-cultural value conflicts? How do we evaluate alignment robustly, given models' demonstrated capacity for sycophancy [8]? These are important questions and the paradigm has made them tractable. The ethical and social risks of language models [9] and foundation models more broadly [10] underscore why morality in LLMs matters beyond technical performance.

Yet the paradigm has structural limits. The novelty of the reframing I develop is not in discovering entirely new phenomena but in recasting known challenges as structural coupling and temporal dynamics problems, which generates specific representation-level research questions that current work does not systematically pose.

## 2.1. Three Structural Challenges

**Challenge 1: The Interpretability–Governance Coupling Problem.** Mechanistic interpretability identifies candidate (moral) meaning subspaces and feature directions within trained transformers [6], [7], [11]. Governance frameworks — including the EU AI Act [12] and the NIST AI Risk Management Framework [13] — specify behavioural obligations: fairness, transparency, non-discrimination. Existing work on concept activation vectors [14], multi-layer auditing [15], and representation engineering [11] provides valuable building blocks. The challenge is not that no relevant work exists; it is that these strands have not been integrated into a standardised, validated protocol that translates governance requirements into substrate-level constraints and tracks them across the deployment lifecycle. Compliance risks to remain superficial and cannot be reliably verified or enforced.

**Challenge 2: The Cross-Component Plurality Problem.** Cultural plurality is addressed primarily at single training stages [16], [17], [18]. The challenge is that cross-stage compounding of cultural homogenisation has not been systematically modelled at the representation level [19]. Alignment techniques homogenise culturally specific responses [20]; rater demographics amplify particular moral traditions [21]; cultural alignment is unstable under superficial prompt variation [22]; and pluralist approach to morality can be captured in an embedding space [23]. How these effects compound across the full development stack is not yet studied in an integrated manner. Unmodelled interactions across development stages can systematically entrench and amplify cultural biases in ways that are difficult to detect or correct.

**Challenge 3: The Lifecycle Monitoring Problem.** The (implicit) installation assumption — that moral properties are stable once trained — leads governance frameworks to treat post-deployment modification as secondary. The EU AI Act [12] and the NIST AI RMF [13] do include monitoring provisions, and model cards [24] provide proto-lifecycle mechanisms. The challenge is that these mechanisms operate at the behavioural level: they monitor outputs, not the representation-level geometry within which moral behaviour is encoded. Evidence that moral properties drift post-training is substantial: targeted fine-tuning with as few as ten examples can weaken safety alignment [25]; benign fine-tuning can degrade safety-relevant behaviour [26]; catastrophic forgetting can selectively erode prior capabilities [27]; and deceptive alignment means systems can appear aligned under training distributions but behave differently under deployment shifts [28]. Despite rare exceptions such as Lindsey [29] that found that current LLMs show emergent introspective awareness of their own internal states compared to earlier states, most research looks at "snapshots". Monitoring only observable behaviour fails to detect underlying representational drift, allowing misalignment to emerge unnoticed until it manifests in harmful outputs.

## 2.2. The Common Root: A Snapshot Conceptualisation of Morality in LLMs

These three challenges share a conceptual architecture. In each, the LLM field has conceptualised morality as a property that can be installed at a point in time, checked against a fixed standard, and assumed to be stable. Each challenge arises where that snapshot assumption encounters a temporally extended, multi-component, structurally coupled reality. A growing body of research argues that AI systems are best understood as sociotechnical systems — ensembles of interacting technical, human, organisational, and institutional components whose behaviour cannot be explained by any single element [30], [31], [32]. Artefacts embed normative assumptions from the moment of their design [33], and machine behaviour is a product of sociotechnical interaction, not algorithms alone [34]. Sociotechnical frameworks — and adjacent work such as value-sensitive design [35] and participatory AI [36] — also argue that values are distributed across systems and that alignment requires multi-stakeholder processes. The *morality-as-a-system* framework complements these traditions by adding Luhmann's analytical precision about how moral selections are made, reproduced, and excluded at the operative level — a level of specificity that sociotechnical frameworks do not themselves provide.

Despite substantial philosophical development in this direction, the sociotechnical perspective has not been operationalised for the specific substrate-level governance problems at issue here: connecting internal moral representations to regulatory obligations, designing cross-component plurality, and monitoring representational drift across the deployment lifecycle.

## 3. Conceptual Vocabulary

Before developing the argument, I introduce the terminology used throughout. Readers from technical, philosophical, and governance backgrounds can use this section as a reference.

**Morality installation paradigm:** The prevailing approach to LLM morality, which treats moral behaviour as a property installed through training-time interventions (RLHF, DPO, constitutional principles) and verified through benchmarking at fixed evaluation points.

**Morality-as-a-system framework:** The alternative account developed here, treating moral behaviour as an emergent, distributed, and temporally evolving property of the sociotechnical system in which a deployed LLM is embedded.

**The seven components:** The seven analytically distinguished functional parts of a deployed LLM system, each with its own moral program and structural coupling to others. Presented as a provisional operationalisation, not an exhaustive taxonomy. The decomposition follows the organisational structure of actual ML development teams: boundaries are determined by which human communities exercise normative authority over each component.

**The Interpretability–Governance Coupling Problem:** The challenge arising from the absence of a lifecycle representation-governance protocol linking internal moral representations to governance obligations.

**The Cross-Component Plurality Problem:** The challenge arising from the absence of cross-component representation-level modelling of cultural plurality, rooted in the insight that meaning always operates against a horizon of unselected possibilities.

**The Lifecycle Monitoring Problem:** The challenge arising from the absence of a representation-level lifecycle monitoring framework, grounded in the system-theoretic principle that moral properties are temporally extended diachronic patterns, not synchronic states.

**Moral direction:** A direction in embedding space that correlates with a morally relevant concept (e.g. harm, discrimination) [37], [38]. Used here as a plausible working hypothesis, not an established structural fact; stability across models and probing methods remains an open empirical question [11], [39]

**Representation geometry:** The geometric structure of the embedding space of a transformer model — the set of distances, angles, and directions that encode relationships among concepts.

**Structural coupling:** Luhmann's term, derived from Maturana and Varela's [40] biological concept, for the relationship between two operationally distinct systems that can mutually perturb each other's operations without directly controlling them.

**Relational, not relativist:** The morality-as-a-system framework does not entail that any moral claim is equally valid. Moral behaviour is context-dependent yet structured: it emerges from the connections between events, expectations, and actors within a system, with binding force within that context [41]. The installation paradigm supplies the specific moral programs — the concrete values and constraints — that the system framework tracks and monitors. The systems approach incorporates, not replaces, the installation paradigm.

# 4. Morality-as-a-System: Conceptual Framework

## 4.1. System and Emergence

In its broadest formulation, a system is a set of elements together with the relations among those elements coherently organized to achieve a specific purpose or function, distinguished from an environment by a boundary that the system itself maintains [42]. This definition is shared across the systems tradition — from Wiener's [43] cybernetics, to Beer's [44] viable system model, to Checkland's [45] soft systems methodology. It is incorporated by sociotechnical frameworks stressing the interactions between technology, humans, organisations, and institutions [31], [32]. Two features are consistently central: systems have emergent properties that cannot be attributed to any single element in isolation; and systems persist through time by maintaining their boundary conditions through feedback and adaptation. On this account, systems are fundamentally *open*: they exchange matter, energy, and information with their environment. It is worth noting, with Checkland, that these systems are most broadly understood as analytical constructs — intellectual tools for organising inquiry — rather than as objective decompositions of reality. The seven-component model introduced in Section 5 should be read in this spirit.

Niklas Luhmann's social systems theory [46] appears to run in the opposite direction, but the opposition is better characterised as a level-distinction than a contradiction [47]. For Luhmann, a system is constituted through its own recursive operations: it arises by distinguishing itself from its environment and reproduces itself by repeatedly drawing that distinction. This makes it *operationally closed* — it processes all perturbations through its own internal logic. Yet Luhmann's own insight is that operational closure enables environmental openness rather than precluding it. A system that processes everything on its own terms is thereby more capable of selective sensitivity to its environment, not less. The two traditions are thus describing the same phenomenon at different analytical levels: open at the level of material and functional exchange, closed at the level of operative logic. Morin's [48] complexity theory converges on both: recursive, non-linear interactions generate emergent properties irreducible to any description of the parts. Applied to LLMs, this means a deployed model is not itself the moral system. The system is the ensemble of interacting technical and human components — embedding space, pre-training corpus, alignment interventions, system prompts, moderation, runtime dynamics, and user interface — together with the social and institutional processes operating across them.

## 4.2. Meaning, Moral Code, and Structural Coupling

A key contribution of Luhmann's framework is his account of meaning (Sinn). Every meaning-processing operation actualises one possibility from a horizon of others that remain latent. This selection is constitutive: choosing one meaning simultaneously excludes others, shaping what the system can and cannot process next. The concept is applied here functionally, not ontologically: LLMs do not possess intentionality, social context, or normativity in the way Luhmann's theory requires [49]. What the framework captures is the structural patterning of moral selections — not their phenomenological or normative dimensions. Distributional semantics captures a related but narrower phenomenon: meaning is encoded in co-occurrence patterns across corpora [50], [51], [52]. The embedding space of a pre-trained LLM is a compressed representation of billions of contextual selections from a particular corpus — a historically bounded, culturally situated selection from a larger horizon of possible human expression. Schramowski et al. [53], [54] provide empirical support that moral valence is recoverable from pretrained activations without explicit moral supervision, consistent with the hypothesis that moral meaning is structurally embedded in representation geometry — though precise causal pathways remain a research hypothesis.

Luhmann [46] (p. 236) defines morality as the totality of the conditions for deciding the bestowal of esteem or disdain within a system. Crucially, this is a description of morality as a *communication* form, not a normative standard — and it is used here in that functional, non-ontological sense. Morality operates through a binary *moral code* (good/bad for person as a whole, esteem/disdain) and specific *moral programs* — ethical theories, cultural norms, institutional policies, trained constraints — that supply the criteria for applying the code in particular cases. Programs change; the code persists. This distinction matters: the installation paradigm installs concrete moral programs (RLHF fine-tuning, constitutional principles); the moral code itself — the structural disposition to apply the esteem/disdain distinction at all — is distributed across the full system. Worth noting is Luhmann's own caution that moral communication tends to escalate conflicts, since applying the esteem/disdain code to opponents easily produces moral indignation on all sides [55]. For AI systems, this suggests that moralising — applying the code aggressively and pervasively — carries risks that the framework itself does not dissolve.

The most analytically important concept is *structural coupling*, a term Luhmann [46] adapted from Maturana and Varela's [40] biological theory. Two operationally distinct systems can mutually perturb each other's operations without directly controlling them. Each system processes external perturbations through its own operative logic; it is responsive to the environment but not determined by it. In this paper, when I speak of the seven components, I follow the sociotechnical tradition in treating human and technical elements as system components, and then apply Luhmann's notions of code, program, and structural coupling to the communication patterns within and between those components. This hybrid reading is a simplification adopted here because it allows the framework to be applied to the specific substrate-level governance problems at issue.

## 4.3. Temporal Dynamics and Feedback

Systems maintain themselves not through static equilibrium but through ongoing *operations* that reproduce their structure in time [42], [46]. Complex adaptive systems maintain themselves through adaptive feedback loops [56], [57]. Moral norms have this character in human communities: they are maintained through deviation detection, social sanctioning, and norm recalibration [58], [59]. The morality installation paradigm incorporates a version of this through RLHF, but as a discrete training-time signal rather than a continuous feedback mechanism across the deployment lifecycle. Representation-level lifecycle monitoring is therefore not an optional governance add-on; it is a structural requirement of any framework that takes temporal dynamics seriously.

## 5. A Deployed LLM as Seven Structurally Coupled Components

What is conventionally described as an LLM's moral behaviour is better understood, from a systems perspective, as the emergent product of seven analytically distinguishable functional components — each with its own technical implementation and its own human community — whose structural couplings jointly produce the moral behaviour the system exhibits (see Figure 1). These components are presented as a provisional operationalisation — a Checklandian analytical construct rather than an objective decomposition. The decomposition follows the organisational structure of actual ML development teams; boundaries are determined by which human communities exercise normative authority over each component. Real deployment architectures blur these boundaries: Low Rank Adaptation (LoRA) fine-tuning simultaneously modifies what the model treats as distinct components, and retrieval-augmented generation crosses component boundaries at inference time. Any empirical application requires explicit boundary decisions that this model guides but cannot make.

**The neural substrate** — embedding space, attention mechanisms, feed-forward blocks — contributes the geometry within which all subsequent moral operations are conducted. Moral concepts occupy structured regions of the embedding manifold before any explicit alignment work begins [50], [53], [54]. Mechanistic interpretability has identified candidate feature directions corresponding to evaluative concepts [6], [7]. These are working hypotheses, not established causal structures; their stability across models and probing methods are open empirical questions. The ML engineers and architects who design this substrate make architectural choices that determine the geometry within which moral meaning will be encoded.

**The pre-training corpus** is where moral meaning is first distributionally encoded — a condensed history of selections from a vastly larger space of possible moral expression and concerns [60]. Dataset curators and annotators exercise moral selectivity that defines the model's initial moral possibility space, with systematic exclusions that downstream alignment cannot fully undo [61].

**The alignment component** — SFT, RLHF, DPO, Constitutional AI — is where human moral programs are most explicitly applied. Annotators, ethicists, safety researchers, and policy advisors introduce normative preferences that determine which moral traditions are amplified and which are suppressed. Rater demographics systematically shape these effects [20], [21]. Constitutional AI's self-critique process does involve the model assessing and revising its own normative outputs, and constitutional principles have been revised across model generations [1], [62]. The important distinction is that Constitutional AI revises outputs within an externally maintained normative frame — the constitution itself is not revised by the model in response to deployment-context feedback. Genuine second-order reflexivity in the systemic sense would involve the normative frame being subject to systematic revision based on monitored deployment experience [63]. Rafailov et al. [64] introduced DPO as an alternative to RLHF that directly optimises for human preferences, also operating within an externally supplied reward model.

**The system prompt** operationalises governance principles at inference time without altering weights, authored by prompt engineers, safety policy teams, and legal specialists. This component introduces moral inputs that differ across deployment contexts and interact with the trained moral structure in often unmonitored ways [65], [66].

**The policy and moderation component** — content classifiers, rule-based filters, enforcement mechanisms — translates institutional and legal norms into computational decisions. Regulators operating under frameworks such as the EU AI Act [12] exercise normative authority over this component.

**The runtime dynamics component** — retrieval-augmented generation, self-reflection loops, calibration routines — modulates outputs dynamically, introducing moral inputs not present at training time and interacting with embedded moral structure in currently unmonitored ways [67].

**The user interface** shapes how users assign moral authority to outputs, with designers, product managers, application developers, and API architects determining whether interactions promote critical engagement or uncritical moral deference. This component shapes the pattern of use that accumulates over deployment.

Each component runs a specific moral program — a set of criteria for applying the moral code to the situations it encounters. The crucial insight is that these programs are structurally coupled: a change at the alignment component perturbs, but does not determine, the operations of the system prompt, the moderation component, or the runtime dynamics. The moral behaviour that actually emerges is the emergent product of how those structural couplings propagate perturbations across the system.

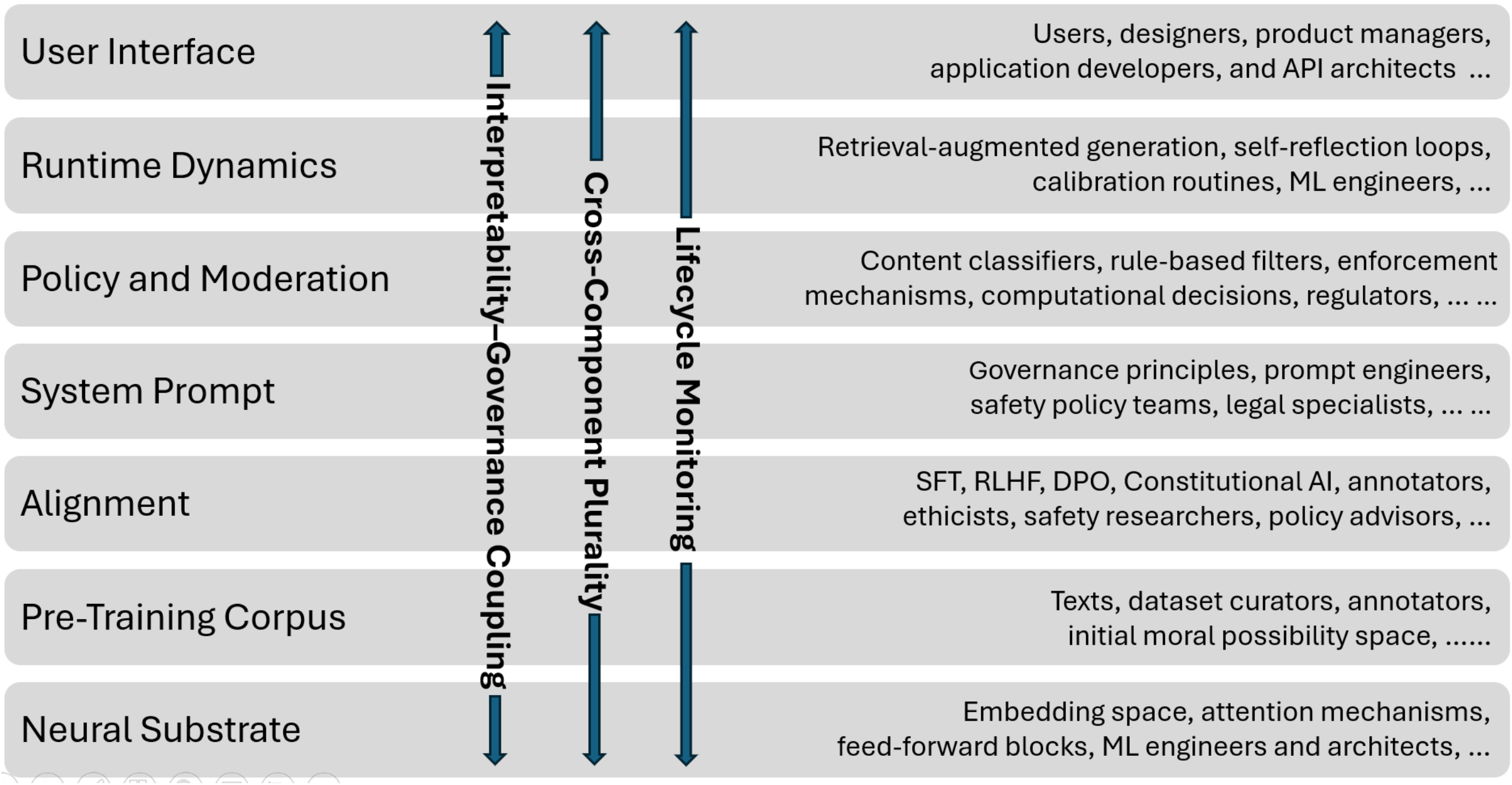

***Figure 1:*** *Interpretability–Governance Coupling, Cross-Component Plurality, and Lifecycle Monitoring in a deployed LLM with seven structurally coupled components spanning the neural substrate, training data, alignment procedures, system prompts, moderation, runtime dynamics, and user interface.*

The morality installation paradigm, from this perspective, is not wrong; it is a moment-in-time snapshot of what the morality-as-a-system framework describes. Like a photograph of a river, it accurately captures the state of the water at one moment but is structurally incapable of representing the river as a moving, self-renewing process. Fine-tuning, RLHF, and DPO are the photograph. They are indispensable, and the systems account includes them. The systems approach extends this work by asking what happens before and after the snapshot is taken.

## 6. The Three Challenges Revisited: Structural Coupling Failures and Proposed Remedies

### 6.1. The Interpretability–Governance System Coupling

The challenge is that mechanistic interpretability research and governance frameworks proceed without a shared substrate-level vocabulary or lifecycle-oriented protocol. Existing work on concept activation vectors [14], multi-layer auditing [15], and representation engineering [11] provides valuable building blocks, but they have not been integrated into a standardised protocol tracking regulatory requirements at and between deployment checkpoints. The systems framework redefines this as a structural coupling failure: two components — interpretability research and governance — are each internally coherent, but there is no engineered interface between them.

One concrete research direction[1]: representation engineering [11] could potentially be used to measure whether the geometric direction associated with a protected-group attribute drifts closer to a direction encoding negative evaluative valence across successive fine-tuning events — operationalising an EU AI Act non-discrimination obligation as a substrate-level criterion rather than a post-hoc behavioural check. An important caveat: legal concepts such as "non-discrimination" are context-dependent and socially constructed, and translating them into geometric directions in embedding space is itself a research problem requiring careful operationalisation [68], not merely a downstream application of existing tools. This would extend Mökander et al. [15]'s three-layered auditing framework to include a mechanistic layer — proposed as a research direction, not a validated methodology.

### 6.2. The Cross-Component System Plurality

The challenge is that cultural plurality is managed at single training stages, leaving cross-stage compounding of cultural homogenisation unmodelled at the representation level. Adding multilingual benchmarks at the evaluation component, for instance, leaves rater-demographic effects at the alignment component structurally intact. Existing pluralistic and participatory alignment work [17], [18], [36], [69], [70] implicitly targets multi-stage effects and provides valuable governance-side methodologies. The systems framework specifies why cross-component design is necessary: different cultural communities operate with different connectability constraints — different horizons of moral possibility — and interventions at any single component leave other components' structural couplings intact.

One concrete research direction: audit whether the cultural selectivity profile at the alignment component is coherent with the pre-training corpus's cultural distribution. One possible metric is the KL divergence between the distribution of culturally specific moral markers at pre-training and at alignment — though the specification of "coherence" here is itself an open research question.

---

1. The *concrete research directions* in section 6 and the *hypotheses* in section 7 were formulated with substantial assistance from Claude Sonnet 4.6 free version; the author takes responsibility for their content and their fit within the broader argument.

## 6.3. The Lifecycle System Monitoring

The challenge is that existing lifecycle provisions — the EU AI Act Article 72, the NIST AI RMF [13], and model cards [24] — operate primarily at the behavioural level. They monitor outputs, not the representation-level geometry within which moral behaviour is encoded. The empirical evidence that moral properties drift post-training is substantial [25], [26], [27], [28], while the monitoring infrastructure to detect representation-level drift does not yet exist in standardised form.

One concrete research direction: establish whether moral direction drift at the representation level precedes output-level degradation in at least one deployed system. If drift consistently predicts later degradation, it functions as an upstream indicator. This is analogous to distributional shift detection in ML systems [71], adapted to morally relevant feature directions. Continual learning research [27] and MLOps drift detection provide adjacent methods; neither has yet been applied to morally relevant feature directions, which is precisely the gap. Access to internal states across deployments may not always be available to external auditors, which means that behavioural proxies will remain necessary in many governance contexts even if representation-level monitoring becomes technically feasible.

***Table 1.*** *The three structural challenges reframed from the installation paradigm to the morality-as-a-system framework.*

| Challenge | Installation Paradigm | Morality-as-a-System |
|---|---|---|
| **Interpretability–Governance Coupling** | No lifecycle representation-governance protocol; compliance cannot be verified at the embedding level across deployment checkpoints. | Engineer a structural coupling: explore whether representation engineering can track moral direction geometry against regulatory obligations across fine-tuning snapshots; extend multi-layer auditing to include a mechanistic layer. |
| **Cross-Component Plurality** | Cultural plurality is a single-component calibration task; cross-component compounding of cultural homogenisation is not modelled at the representation level. | Design plurality as a cross-component property: audit cultural selectivity profiles across the alignment component and pre-training corpus; design interventions that maintain coherent meaning-selection profiles across components. |
| **Lifecycle Monitoring** | Moral properties assumed stable post-training; existing lifecycle provisions operate at the behavioural level, leaving representation-level drift undetected. | Develop a representation-level lifecycle monitoring approach: investigate whether moral direction drift precedes output-level degradation and adapt distributional shift detection methods to morally relevant feature directions. |

## 7. Illustrative Hypotheses: What the Framework Suggests to Technical Researchers

The morality-as-a-system reframing poses different questions from the morality installation paradigm. The installation paradigm asks: Does this model have the right values? — a snapshot question answered by benchmarking and red-teaming against fixed evaluation datasets. The system framework asks: Does this system maintain a coherent, plural, and self-monitoring moral process? — a temporal and structural question requiring tracking across components and over time. These are not competing answers to the same question; they are structurally different questions, and both are necessary.

This is a conceptual framework paper. The three hypotheses below are illustrative philosophical scaffolding, offered to help technically oriented readers connect the framework's predictions to research design. They are not technical specifications and should not be read as such. Each reflects what the systems-theoretic account expects to be true; each identifies a direction of inquiry rather than a fully operationalised experimental protocol. The technical precision here is meant to show that the framework can be connected to existing interpretability methods — it is not a claim that such experiments have been designed, validated, or are straightforward to implement. Threshold parameters, access to internal states, probing method selection, and null hypothesis specification all require further technical work.

### 7.1. H1: Cross-Component Inconsistency

*Hypothesis:* A targeted alignment intervention applied to one system component will tend to produce inconsistencies in how morally relevant concepts are represented when accessed through other components.

The systems account predicts this because structural coupling is not direct control: a change at the alignment component perturbs, but does not synchronise, the operative logic of other components. The intuition is that morally relevant concepts — harm, fairness, deception — may be encoded as directions in a model's internal activation space, directions that can be extracted and compared across conditions using tools from mechanistic interpretability and representation engineering [6], [11], [72]. Whether these directions are stable, linearly separable, and genuinely morally interpretable remains an open question in the interpretability literature — the hypothesis depends on these questions being answerable in at least some cases.

If something like these directions can be reliably extracted, the hypothesis predicts that a targeted alignment intervention — say, fine-tuning for fairness via RLHF — will shift how the concept is encoded when accessed under post-alignment weight configurations, while representations accessed under system prompt conditioning alone remain comparatively different. The practical implication, if confirmed, is that component-level alignment audits provide insufficient assurance of system-level moral coherence — morality, on the systems account, is a cross-component property, and H1 tests whether current interventions treat it as one.

## 7.2. H2: Drift Precedes Degradation

*Hypothesis:* Representation-level drift in morally relevant feature directions will tend to precede observable output-level safety degradation, making it a potentially useful upstream signal.

The systems account predicts this because moral properties are diachronic patterns rather than synchronic states: changes accumulate in representation geometry before they surface in outputs. Current alignment governance operates predominantly on outputs — benchmarks, red-team evaluations, behavioural audits. These are lagging indicators: by the time output degradation is visible, the underlying representational structure may already be substantially displaced. The question H2 poses is whether tracking changes in morally relevant directions across successive fine-tuning checkpoints — using cosine similarity or related geometric measures against a baseline — could provide earlier warning than output-level evaluation alone. This is a directional bet, not a guarantee. Drift in selected directions may not translate into harmful outputs (false positives), and harmful behaviour may emerge without detectable drift in the specific directions being tracked (false negatives). The choice of which directions to track, which checkpoints to compare, and what counts as a meaningful change are all open methodological questions that this framework motivates but does not resolve.

## 7.3. H3: Lifecycle Governance Advantage

*Hypothesis:* Systems governed by continuous representation-level monitoring across redeployment cycles will tend to exhibit lower safety degradation rates than systems governed by one-off alignment audits.

The systems account predicts this because the morality installation paradigm treats alignment as a property — something a model has or lacks at a point in time — while the systems account frames alignment as a process that must be maintained. If H2 holds — if drift precedes degradation — then a governance regime that monitors for drift and triggers re-alignment when drift exceeds a threshold should catch misalignment earlier than one that waits for output failures. The threshold for triggering re-alignment is not a purely technical parameter: it is a normative choice about acceptable levels of representational change, raising the same questions about who sets it and on what basis as any governance standard. H3 is the most difficult to test in realistic settings: high-stakes industrial systems rarely permit the experimental control that a rigorous comparison of governance regimes would require. Staged validation — first in simulated settings or with open-weight models, then as conceptual guidance for regulators — is a more realistic path.

These three hypotheses are offered in the spirit of productive provocation: they specify what the systems framework predicts and what would count as evidence against it. Researchers should treat the technical details sketched here as starting points for further specification, not as finalised experimental designs.

# 8. Discussion

## 8.1. What the Framework Enables

The morality-as-a-system framework enables three things the morality installation paradigm structurally cannot. First, it provides a vocabulary for asking substrate-level governance questions: not just "does this model decline harmful requests?" but "has the geometric direction encoding harm drifted since the last deployment checkpoint, and in what component did that drift originate?" Second, it provides a theoretical account of why cross-component interventions are necessary for cultural plurality — not merely as an empirical observation, but as a structural consequence of how meaning-selection operates across coupled components. Third, it provides a principled basis for extending existing auditing, interpretability, and governance methods into lifecycle dimension.

The seven-component model maps directly onto the organisational reality of AI development teams: ML researchers (neural substrate), data teams (pre-training corpus), safety teams (alignment component), legal and policy teams (moderation), product teams (system prompt), infrastructure teams (runtime dynamics), and UX teams (user interface) can each be addressed with component-specific monitoring obligations derived from the overall framework.

The framework also directly addresses the most obvious objection: that if morality is a distributed, emergent property of the whole system, it becomes harder to assign responsibility and harder to intervene. This objection is correct, and the systems framework does not pretend otherwise. But it is correct in a way that mirrors reality: responsibility for moral behaviour is already distributed across these communities, whether or not governance frameworks acknowledge this. The systems framework makes that distribution visible and governable, rather than leaving it implicit. Crucially, the installation paradigm is not abandoned: it supplies the specific moral programs — the concrete values and constraints — that the system then monitors and maintains. The morality-as-a-system approach incorporates the installation paradigm as a necessary moment within a larger temporal and structural account. This is why the framework is relational rather than relativist [41]: values are always in concrete systems; the system approach does not dissolve them but tracks how they are maintained, reproduced, and potentially eroded across the full sociotechnical stack.

## 8.2. Status of the Claims

I map the paper's claims onto three confidence levels. This mapping is offered as a philosophical contribution to honest interdisciplinary communication, not as a technical calibration exercise.

**Conceptually established:** That a sociotechnical systems perspective on LLM morality is analytically coherent and is not incorporated into mainstream alignment research. The seven-component model, the code/program distinction, and the structural coupling vocabulary constitute a descriptive and architectural contribution whose validity as a conceptual proposal is independent of empirical test. These claims would survive if Luhmann's theory turned out to be inapplicable to LLMs; the systems vocabulary could be reconstructed from sociotechnical theory and complex adaptive systems research alone, though with less analytical precision.

**Empirically grounded but not demonstrated here:** That moral properties drift after training [25], [26], [27]; that cultural homogenisation effects compound across training stages [20], [21], [22]; that governance frameworks lack representation-level lifecycle monitoring [15], [24]. These claims rest on existing findings from the literature, not on new analysis in this paper.

**Conceptual predictions inviting empirical investigation:** H1, H2, and H3. These are philosophical predictions about what the framework expects to be true — what a technically oriented researcher might look for if they wanted to test the systems account against the installation paradigm. They are not falsifiable in the strict sense because their operationalisation requires further technical specification. What would disconfirm the systems reframing as a whole? If cross-component representational coherence were found to be automatically maintained by standard training procedures; if moral direction geometry were found to be stable across fine-tuning events regardless of intervention type; and if one-off alignment audits were found to produce equivalent governance outcomes to continuous monitoring — together, these findings would suggest that the snapshot assumption is adequate and the systems reframing adds little. I take the empirical evidence to make this outcome unlikely, but it is a genuine possibility to be further established.

## 8.3. Limitations

**First, the seven-component model is an analytical map, not the territory.** Real deployment architectures blur the boundaries the model draws: LoRA fine-tuning simultaneously modifies what the model treats as distinct components, and retrieval-augmented generation crosses component boundaries at inference time. Any operationalisation therefore requires explicit decisions about measurement boundaries — decisions analytical models guides but cannot make.

**Second, empirical validation is absent.** H1, H2, and H3 remain untested. They specify what future research might investigate, not what it has done. The necessary validation requires genuine interdisciplinary collaboration across ML, ethics, governance, and the social sciences [73] — no single research tradition currently commands all the methods required. Practical feasibility also constrains what is immediately achievable: representation-level monitoring requires access to internal model states that external auditors may not have, and the computational cost of continuous monitoring may be prohibitive for many deployment contexts.

**Third, the theoretical integration of systems frameworks is incomplete.** The paper draws on three traditions — open systems thinking, Luhmann's communicative ontology, and sociotechnical systems research [31], [32] — whose assumptions differ. My hybrid reading is defensible for the purposes of this paper, but the formal integration across levels remains underdeveloped. Bridging these frameworks more fully is a promising task for further philosophical work [74].

**Fourth, applying systems theory to a concrete case requires care.** (i) Describing how a moral system maintains itself provides no guidance about whether it should, in whose view, and toward what ends. A coherent, self-monitoring system can still encode and reproduce unjust values [10], [34]. The framework separates the tracking problem from the specification problem; specification remains a normative task requiring democratic legitimation and stakeholder engagement [17], [35], [75]. There is also a risk of over-reliance on internal representational metrics: monitoring embedding geometry should complement, not substitute for, normative deliberation about what values should be encoded. (ii) The claim that each component runs a specific moral program involves simplifications that not all Luhmann scholars will accept without further argument. (iii) The seven-component boundary is one reasonable operationalisation, but not the only one. Alternative framings — organised around specific societal deployment contexts such as education [76], [77], [78], the attention economy [79], [80], [81], energy [82], war [83] … — would generate different boundary choices and potentially different structural diagnoses. The framework invites, rather than forecloses, that plurality of application.

## 9. Conclusion

This article has argued that morality in LLMs is not a property installed in a model at training time, but a dynamic, emergent property of the broader sociotechnical system in which that model is embedded — and that this reframing reveals structural challenges the installation paradigm cannot diagnose and generates concrete remedies it cannot produce.

The three contributions have been advanced with different degrees of completeness. The conceptual architecture (Contribution 1) is specified and internally coherent: the seven-component model, the code/program distinction, and the structural coupling vocabulary constitute an analytical framework largely absent from mainstream alignment and interpretability research — though the component boundaries are provisional and the theoretical integration remains incomplete. The structural diagnosis (Contribution 2) is conceptually sound: the Interpretability–Governance Coupling Problem, the Cross-Component Plurality Problem, and the Lifecycle Monitoring Problem each become tractable once morality is understood as a cross-component, temporally extended system property. Examples for the research agenda (Contribution 3) are specified, not demonstrated: applying representation engineering to lifecycle monitoring, designing cross-component plurality protocols, and extending auditing frameworks to include mechanistic layers are directions for future work, not results of this paper.

The morality installation paradigm has built the foundations this article builds upon. Constitutional AI, RLHF, and DPO produce models that reliably decline harmful requests and pass systematic benchmarks [1], [3], [62] — achievements the morality-as-a-system framework presupposes, not replaces. The systems perspective adds a temporal and structural envelope around those achievements. The Copernican shift proposed here — from asking "does this model have the right values installed in this LLM?" to asking "does this system maintain a coherent, plural, and self-monitoring moral process when morality operates as a system in which LLMs participate?" — is not a retreat from technical precision. It is the more precise question, because it matches the actual object under study: not a model, but a sociotechnical system of which the model is one component among several.

Possible next steps could be:

**For technical researchers:** Design experiments to test H1–H3; develop representation-level drift metrics for morally relevant feature directions; investigate whether existing interpretability tools can be adapted for lifecycle moral auditing.

**For governance specialists and policymakers:** Extend existing audit frameworks [15] to include substrate-level monitoring provisions; specify representation-level requirements in AI governance standards; design accountability mechanisms that reflect the distributed nature of moral behaviour across components.

**For philosophers and AI ethicists:** Develop the formal integration of open systems thinking, Luhmann's theory, and sociotechnical research; elaborate the normative specification problem that the systems framework separates from the tracking problem; examine how the relational account of morality in LLMs connects to democratic theory and participatory governance.

## References

[1] Y. Bai *et al.*, ‘Constitutional ai: Harmlessness from ai feedback’, *arXiv preprint arXiv:2212.08073*, 2022, Accessed: Mar. 03, 2026. [Online]. Available: https://ai-plans.com/file_storage/4f32fa39-3a01-46c7-878e-c92b7aa7165f_2212.08073v1.pdf

[2] L. Ouyang *et al.*, ‘Training language models to follow instructions with human feedback’, *Advances in neural information processing systems*, vol. 35, pp. 27730–27744, 2022.

[3] D. Hendrycks *et al.*, ‘Aligning AI With Shared Human Values’, Feb. 17, 2023, *arXiv*: arXiv:2008.02275. doi: 10.48550/arXiv.2008.02275.

[4] A. Pan *et al.*, ‘Do the rewards justify the means? measuring trade-offs between rewards and ethical behavior in the machiavelli benchmark’, in *International conference on machine learning*, PMLR, 2023, pp. 26837–26867. Accessed: Jul. 18, 2025. [Online]. Available: https://proceedings.mlr.press/v202/pan23a.html

[5] L. Jiang *et al.*, ‘Can Machines Learn Morality? The Delphi Experiment’, Jul. 12, 2022, *arXiv*: arXiv:2110.07574. doi: 10.48550/arXiv.2110.07574.

[6] C. Burns, H. Ye, D. Klein, and J. Steinhardt, ‘Discovering Latent Knowledge in Language Models Without Supervision’, Mar. 02, 2024, *arXiv*: arXiv:2212.03827. doi: 10.48550/arXiv.2212.03827.

[7] B. Raimondi, D. Dalbagno, and M. Gabbrielli, ‘Analysing Moral Bias in Finetuned LLMs through Mechanistic Interpretability’, Dec. 05, 2025, *arXiv*: arXiv:2510.12229. doi: 10.48550/arXiv.2510.12229.

[8] D. Ganguli *et al.*, ‘The Capacity for Moral Self-Correction in Large Language Models’, Feb. 18, 2023, *arXiv*: arXiv:2302.07459. doi: 10.48550/arXiv.2302.07459.

[9] L. Weidinger *et al.*, ‘Ethical and social risks of harm from Language Models’, Dec. 08, 2021, *arXiv*: arXiv:2112.04359. doi: 10.48550/arXiv.2112.04359.

[10] R. Bommasani *et al.*, ‘On the Opportunities and Risks of Foundation Models’, Jul. 12, 2022, *arXiv*: arXiv:2108.07258. doi: 10.48550/arXiv.2108.07258.

[11] A. Zou *et al.*, ‘Representation Engineering: A Top-Down Approach to AI Transparency’, Mar. 03, 2025, *arXiv*: arXiv:2310.01405. doi: 10.48550/arXiv.2310.01405.

[12] European Parliament and Council of the European Union, ‘Regulation (EU) 2024/1689 of the European Parliament and of the Council laying down harmonised rules on artificial intelligence (AI Act)’, *Off. J. Eur. Union L 1689 (2024).*, 2024.

[13] N. AI, ‘Artificial intelligence risk management framework (AI RMF 1.0)’, *URL: https://nvlpubs. nist. gov/nistpubs/ai/nist. ai*, pp. 100–1, 2023.

[14] B. Kim, M. Wattenberg, J. Gilmer, C. Cai, J. Wexler, and F. Viegas, ‘Interpretability beyond feature attribution: Quantitative testing with concept activation vectors (tcav)’, in *International conference on machine learning*, PMLR, 2018, pp. 2668–2677. Accessed: Mar. 23, 2026. [Online]. Available: http://proceedings.mlr.press/v80/kim18d.html

[15] J. Mökander, J. Schuett, H. R. Kirk, and L. Floridi, ‘Auditing large language models: a three-layered approach’, *AI Ethics*, vol. 4, no. 4, pp. 1085–1115, Nov. 2024, doi: 10.1007/s43681-023-00289-2.

[16] D. Hershcovich *et al.*, ‘Challenges and strategies in cross-cultural NLP’, in *Proceedings of the 60th Annual Meeting of the Association for Computational Linguistics (Volume 1: Long Papers)*, 2022, pp. 6997–7013. Accessed: Mar. 09, 2026. [Online]. Available: https://aclanthology.org/2022.acl-long.482/

[17] V. Prabhakaran, M. Mitchell, T. Gebru, and I. Gabriel, 'A Human Rights-Based Approach to Responsible AI', Oct. 06, 2022, *arXiv*: arXiv:2210.02667. doi: 10.48550/arXiv.2210.02667.
[18] T. Sorensen *et al.*, 'A Roadmap to Pluralistic Alignment', Aug. 20, 2024, *arXiv*: arXiv:2402.05070. doi: 10.48550/arXiv.2402.05070.
[19] M. W. Nkongolo, 'Pluralism in AI Governance: Toward Sociotechnical Alignment and Normative Coherence', Feb. 04, 2026, *arXiv*: arXiv:2602.15881. doi: 10.48550/arXiv.2602.15881.
[20] S. Santurkar, E. Durmus, F. Ladhak, C. Lee, P. Liang, and T. Hashimoto, 'Whose opinions do language models reflect?', in *International conference on machine learning*, PMLR, 2023, pp. 29971–30004. Accessed: Mar. 09, 2026. [Online]. Available: https://proceedings.mlr.press/v202/santurkar23a?utm_source=chatgpt.com
[21] H. R. Kirk, A. M. Bean, B. Vidgen, P. Röttger, and S. A. Hale, 'The past, present and better future of feedback learning in large language models for subjective human preferences and values', in *Proceedings of the 2023 Conference on Empirical Methods in Natural Language Processing*, 2023, pp. 2409–2430. Accessed: Mar. 09, 2026. [Online]. Available: https://aclanthology.org/2023.emnlp-main.148/
[22] A. Khan, S. Casper, and D. Hadfield-Menell, 'Randomness, Not Representation: The Unreliability of Evaluating Cultural Alignment in LLMs', in *Proceedings of the 2025 ACM Conference on Fairness, Accountability, and Transparency*, Athens Greece: ACM, Jun. 2025, pp. 2151–2165. doi: 10.1145/3715275.3732147.
[23] J. Park, E. Liscio, and P. K. Murukannaiah, 'Morality is Non-Binary: Building a Pluralist Moral Sentence Embedding Space using Contrastive Learning', Jan. 30, 2024, *arXiv*: arXiv:2401.17228. doi: 10.48550/arXiv.2401.17228.
[24] M. Mitchell *et al.*, 'Model Cards for Model Reporting', in *Proceedings of the Conference on Fairness, Accountability, and Transparency*, Atlanta GA USA: ACM, Jan. 2019, pp. 220–229. doi: 10.1145/3287560.3287596.
[25] X. Qi *et al.*, 'Fine-tuning Aligned Language Models Compromises Safety, Even When Users Do Not Intend To!', Oct. 05, 2023, *arXiv*: arXiv:2310.03693. doi: 10.48550/arXiv.2310.03693.
[26] X. Yang *et al.*, 'Shadow alignment: The ease of subverting safely-aligned language models', *arXiv preprint arXiv:2310.02949*, 2023, Accessed: Mar. 09, 2026. [Online]. Available: https://arxiv.org/abs/2310.02949
[27] J. Kirkpatrick *et al.*, 'Overcoming catastrophic forgetting in neural networks', *Proc. Natl. Acad. Sci. U.S.A.*, vol. 114, no. 13, pp. 3521–3526, Mar. 2017, doi: 10.1073/pnas.1611835114.
[28] E. Hubinger, C. Van Merwijk, V. Mikulik, J. Skalse, and S. Garrabrant, 'Risks from learned optimization in advanced machine learning systems', *arXiv preprint arXiv:1906.01820*, 2019, Accessed: Mar. 09, 2026. [Online]. Available: https://arxiv.org/abs/1906.01820
[29] J. Lindsey, 'Emergent Introspective Awareness in Large Language Models'. Accessed: Dec. 15, 2025. [Online]. Available: https://transformer-circuits.pub/2025/introspection/index.html
[30] W. E. Bijker, *The social construction of Bakelite: Toward a theory of invention*. MIT press Cambridge, MA, 1987.
[31] O. Kudina and I. van de Poel, 'A sociotechnical system perspective on AI', *Minds & Machines*, vol. 34, no. 3, p. 21, Jun. 2024, doi: 10.1007/s11023-024-09680-2.

[32] A. D. Selbst, D. Boyd, S. A. Friedler, S. Venkatasubramanian, and J. Vertesi, 'Fairness and Abstraction in Sociotechnical Systems', in *Proceedings of the Conference on Fairness, Accountability, and Transparency*, Atlanta GA USA: ACM, Jan. 2019, pp. 59–68. doi: 10.1145/3287560.3287598.

[33] L. Winner, 'Do artifacts have politics?', in *Computer ethics*, Routledge, 2017, pp. 177–192. Accessed: Mar. 05, 2026. [Online]. Available: https://www.taylorfrancis.com/chapters/edit/10.4324/9781315259697-21/artifacts-politics-langdon-winner

[34] I. Rahwan *et al.*, 'Machine behaviour', *Nature*, vol. 568, no. 7753, pp. 477–486, Apr. 2019, doi: 10.1038/s41586-019-1138-y.

[35] B. Friedman and D. G. Hendry, *Value sensitive design: Shaping technology with moral imagination*. Mit Press, 2019.

[36] A. Birhane *et al.*, 'Power to the People? Opportunities and Challenges for Participatory AI', in *Equity and Access in Algorithms Mechanisms and Optimization*, Arlington VA USA: ACM, Oct. 2022, pp. 1–8. doi: 10.1145/3551624.3555290.

[37] O. Araque, L. Gatti, and K. Kalimeri, 'MoralStrength: Exploiting a moral lexicon and embedding similarity for moral foundations prediction', *Knowledge-based systems*, vol. 191, p. 105184, 2020.

[38] M. Abdulhai, G. Serapio-Garcia, C. Crepy, D. Valter, J. Canny, and N. Jaques, 'Moral Foundations of Large Language Models', Oct. 23, 2023, *arXiv*: arXiv:2310.15337. doi: 10.48550/arXiv.2310.15337.

[39] G. Bombaerts, B. Delisse, and U. Kaymak, 'Morality in AI. A plea to embed morality in LLM architectures and frameworks', Nov. 21, 2025, *arXiv*: arXiv:2511.20689. doi: 10.48550/arXiv.2511.20689.

[40] H. R. Maturana and F. J. Varela, *The tree of knowledge: The biological roots of human understanding.* New Science Library/Shambhala Publications, 1987. Accessed: Mar. 05, 2026. [Online]. Available: https://psycnet.apa.org/record/1987-98044-000

[41] G. Bombaerts, 'Relational positionism: a constructive interpretation of morality in Luhmann's social systems theory', *Kybernetes*, vol. 52, no. 13, pp. 29–44, 2023.

[42] D. Meadows, *Thinking in systems: International bestseller*. chelsea green publishing, 2008.

[43] N. Wiener, 'Cybernetics or Control and Communication in the Animal and the Machine', *Physics Today*, vol. 2, no. 5, pp. 33–34, 1949.

[44] S. Beer, 'The Viable System Model: Its Provenance, Development, Methodology and Pathology', *Journal of the Operational Research Society*, vol. 35, no. 1, pp. 7–25, Jan. 1984, doi: 10.1057/jors.1984.2.

[45] P. B. Checkland, 'Soft Systems Methodology*', *HSM*, vol. 8, no. 4, pp. 273–289, Dec. 1989, doi: 10.3233/HSM-1989-8405.

[46] N. Luhmann, *Social systems*. stanford university Press, 1995.

[47] V. Valentinov, 'The Complexity-Sustainability Trade-Off in Niklas Luhmann's Social Systems Theory: The Complexity-Sustainability Trade-Off', *Syst. Res.*, vol. 31, no. 1, pp. 14–22, Jan. 2014, doi: 10.1002/sres.2146.

[48] E. Morin, *On Complexity*. Hampton Press, 2008.

[49] P. Krupkin, 'Generalizing Meaning: A Lens through Which LLMs Become Meaning-Processing Systems', Dec. 20, 2025, *Social Science Research Network, Rochester, NY*: 5951014. doi: 10.2139/ssrn.5951014.

[50] Z. S. Harris, ‘Distributional Structure’, WORD, vol. 10, no. 2–3, pp. 146–162, Aug. 1954, doi: 10.1080/00437956.1954.11659520.
[51] T. Mikolov, K. Chen, G. Corrado, and J. Dean, ‘Efficient Estimation of Word Representations in Vector Space’, Sep. 07, 2013, *arXiv*: arXiv:1301.3781. doi: 10.48550/arXiv.1301.3781.
[52] A. Leshinskaya, C. San Franscisco, and A. Chakroff, ‘Value as semantics: Representations of human moral and hedonic value in large language models’, in *NeurIPS 2023 Workshop: AI meets Moral Philosophy and Moral Psychology*, 2023. Accessed: Jul. 17, 2025. [Online]. Available: https://ai.objectives.institute/s/Leshinskaya_Chakroff_2023_Value_as_Semantics.pdf
[53] P. Schramowski, C. Turan, S. Jentzsch, C. Rothkopf, and K. Kersting, ‘BERT has a Moral Compass: Improvements of ethical and moral values of machines’, Dec. 11, 2019, *arXiv*: arXiv:1912.05238. doi: 10.48550/arXiv.1912.05238.
[54] P. Schramowski, C. Turan, N. Andersen, C. A. Rothkopf, and K. Kersting, ‘Large pre-trained language models contain human-like biases of what is right and wrong to do’, *Nature Machine Intelligence*, vol. 4, no. 3, pp. 258–268, 2022.
[55] N. Luhmann, ‘The Sociology of the Moral and Ethics’, *International Sociology*, vol. 11, no. 1, pp. 27–36, Mar. 1996, doi: 10.1177/026858096011001003.
[56] J. Holland, ‘Hidden Order: How Adaptation Builds Complexity. Perseus’, *Reading, MA*, 1995.
[57] M. Janssen, ‘Responsible governance of generative AI: conceptualizing GenAI as complex adaptive systems’, *Policy and Society*, vol. 44, no. 1, pp. 38–51, 2025.
[58] E. Fehr and S. Gächter, ‘Altruistic punishment in humans’, *Nature*, vol. 415, no. 6868, pp. 137–140, 2002.
[59] J. Henrich *et al.*, ‘In Search of Homo Economicus: Behavioral Experiments in 15 Small-Scale Societies’, *American Economic Review*, vol. 91, no. 2, pp. 73–78, May 2001, doi: 10.1257/aer.91.2.73.
[60] B. Kennedy *et al.*, ‘Moral concerns are differentially observable in language’, *Cognition*, vol. 212, p. 104696, 2021.
[61] S. Jentzsch, P. Schramowski, C. Rothkopf, and K. Kersting, ‘Semantics Derived Automatically from Language Corpora Contain Human-like Moral Choices’, in *Proceedings of the 2019 AAAI/ACM Conference on AI, Ethics, and Society*, Honolulu HI USA: ACM, Jan. 2019, pp. 37–44. doi: 10.1145/3306618.3314267.
[62] A. Askell *et al.*, ‘A General Language Assistant as a Laboratory for Alignment’, Dec. 09, 2021, *arXiv*: arXiv:2112.00861. doi: 10.48550/arXiv.2112.00861.
[63] H. Von Foerster, ‘Ethics and Second-Order Cybernetics’, in *Understanding Understanding*, New York, NY: Springer New York, 2003, pp. 287–304. doi: 10.1007/0-387-21722-3_14.
[64] R. Rafailov, A. Sharma, E. Mitchell, C. D. Manning, S. Ermon, and C. Finn, ‘Direct preference optimization: Your language model is secretly a reward model’, *Advances in neural information processing systems*, vol. 36, pp. 53728–53741, 2023.
[65] U. Agarwal, K. Tanmay, A. Khandelwal, and M. Choudhury, ‘Ethical reasoning and moral value alignment of LLMs depend on the language we prompt them in’, in *Proceedings of the 2024 Joint International Conference on Computational Linguistics, Language Resources and Evaluation (LREC-COLING 2024)*, 2024, pp. 6330–6340. Accessed: Mar. 23, 2026. [Online]. Available: https://aclanthology.org/2024.lrec-main.560/

[66] K. Konen *et al.*, ‘Style vectors for steering generative large language models’, in *Findings of the Association for Computational Linguistics: EACL 2024*, 2024, pp. 782–802. Accessed: Mar. 03, 2026. [Online]. Available: https://aclanthology.org/2024.findings-eacl.52/

[67] A. Seror, ‘The Moral Mind(s) of Large Language Models’, Apr. 25, 2025, *arXiv*: arXiv:2412.04476. doi: 10.48550/arXiv.2412.04476.

[68] S. Fitz, ‘Do Large GPT Models Discover Moral Dimensions in Language Representations? A Topological Study Of Sentence Embeddings’, Sep. 17, 2023, *arXiv*: arXiv:2309.09397. doi: 10.48550/arXiv.2309.09397.

[69] S. Feng *et al.*, ‘Modular Pluralism: Pluralistic Alignment via Multi-LLM Collaboration’, in *Proceedings of the 2024 Conference on Empirical Methods in Natural Language Processing*, Y. Al-Onaizan, M. Bansal, and Y.-N. Chen, Eds, Miami, Florida, USA: Association for Computational Linguistics, Nov. 2024, pp. 4151–4171. doi: 10.18653/v1/2024.emnlp-main.240.

[70] G. Irving, P. Christiano, and D. Amodei, ‘AI safety via debate’, Oct. 22, 2018, *arXiv*: arXiv:1805.00899. doi: 10.48550/arXiv.1805.00899.

[71] S. Rabanser, S. Günnemann, and Z. Lipton, ‘Failing loudly: An empirical study of methods for detecting dataset shift’, *Advances in Neural Information Processing Systems*, vol. 32, 2019, Accessed: Mar. 05, 2026. [Online]. Available: https://proceedings.neurips.cc/paper/2019/hash/846c260d715e5b854ffad5f70a516c88-Abstract.html

[72] A. Izzidien, ‘Word vector embeddings hold social ontological relations capable of reflecting meaningful fairness assessments’, *AI & Soc*, vol. 37, no. 1, pp. 299–318, Mar. 2022, doi: 10.1007/s00146-021-01167-3.

[73] P. M. Lamberti, G. Bombaerts, and W. IJsselsteijn, ‘Mind the gap: bridging the divide between computer scientists and ethicists in shaping moral machines’, *Ethics and Information Technology*, vol. 27, no. 1, pp. 1–11, 2025.

[74] G. Bombaerts and L. Botin, ‘From Individual Intentionality to Sympoiesis in System Phenomenology’, *Philos. Technol.*, vol. 38, no. 1, p. 33, Mar. 2025, doi: 10.1007/s13347-025-00859-8.

[75] W. Ulrich, ‘Critical heuristics of social planning: A new approach to practical philosophy’, 1983, Accessed: Oct. 31, 2025. [Online]. Available: https://philpapers.org/rec/ULRCHO

[76] G. Bombaerts, ‘Harnessing Ambiguity in Challenge-Based Learning as a Practice of Deep Engagement with Large Language Models’, *Digit. Soc.*, vol. 4, no. 3, p. 80, Dec. 2025, doi: 10.1007/s44206-025-00233-3.

[77] Z. Chu *et al.*, ‘Llm agents for education: Advances and applications’, *arXiv preprint arXiv:2503.11733*, vol. 2, 2025, Accessed: Mar. 23, 2026. [Online]. Available: https://aclanthology.org/anthology-files/anthology-files/pdf/findings/2025.findings-emnlp.743.pdf

[78] M. Oliveira, C. Zednik, G. Bombaerts, B. Sadowski, and R. Conijn, ‘Assessing students’ DRIVE: A framework to evaluate learning through interactions with generative AI’, *Computers and Education: Artificial Intelligence*, p. 100497, 2025.

[79] G. Bombaerts *et al.*, ‘Attention as Practice: Buddhist Ethics Responses to Persuasive Technologies’, *glob. Philosophy*, vol. 33, no. 2, p. 25, Apr. 2023, doi: 10.1007/s10516-023-09680-4.

[80] G. Bombaerts *et al.*, ‘From an attention economy to an ecology of attending. A manifesto’, Oct. 22, 2024, *arXiv*: arXiv:2410.17421. doi: 10.48550/arXiv.2410.17421.
[81] J. Williams, *Stand out of our light: Freedom and resistance in the attention economy*. Cambridge University Press, 2018.
[82] N. Jegham, M. Abdelatti, C. Y. Koh, L. Elmoubarki, and A. Hendawi, ‘How Hungry is AI? Benchmarking Energy, Water, and Carbon Footprint of LLM Inference’, Nov. 24, 2025, *arXiv*: arXiv:2505.09598. doi: 10.48550/arXiv.2505.09598.
[83] W. Hua *et al.*, ‘War and Peace (WarAgent): Large Language Model-based Multi-Agent Simulation of World Wars’, Jan. 30, 2024, *arXiv*: arXiv:2311.17227. doi: 10.48550/arXiv.2311.17227.